# Real-time Dual-channel 2 × 2 MIMO Fiber-THz-Fiber Seamless Integration System at 385 GHz and 435 GHz

Jiao Zhang[1,2], Min Zhu[1,2],*, Bingchang Hua[2], Mingzheng Lei[2], Yuancheng Cai[1,2], Liang Tian[2], Yucong Zou[2], Like Ma[4], Yongming Huang[1,2], Jianjun Yu[2,3], Xiaohu You[1,2]*

[1] National Mobile Communications Research Laboratory, Southeast University, Nanjing 210096, China
minzhu@seu.edu.cn; xhyu@seu.edu.cn
[2] Purple Mountain Laboratories, Nanjing, Jiangsu 211111, China
[3] Fudan University, Shanghai, 200433, China
[4] Future Mobile Technology Lab, China Mobile Research Institute, Beijing 100053, China

**Abstract** *We demonstrate the first practical real-time dual-channel fiber-THz-fiber 2 × 2 MIMO seamless integration system with a record net data rate of 2 × 103.125 Gb/s at 385 GHz and 435 GHz over two spans of 20 km SSMF and 3 m wireless link.* ©2022 The Author(s)

## Introduction

The Terahertz-band (0.3 THz to 10 THz) is envisioned as a promising candidate for future 6G communication, which can provide hundreds of Gbps or even Tbps data capacity. Photonic heterodyne-based optoelectronic solutions are currently being widely studied due to the large bandwidth and small harmonic interference [1-5]. Moreover, it can facilitate seamless integration with high-speed optical fiber access networks.

Several seamless integration systems of THz wireless and optical fiber networks have been demonstrated enabled by on photonics [6-10]. At 288.5 GHz, an optical-wireless-optical link using an ultra-broadband silicon-plasmonic modulator has been demonstrated with 50 Gb/s line rates over a distance of 16 m [6]. By using a plasmonic modulator with a low noise built-in amplification, the transparent optical-THz-optical link providing line-rates up to 240 and 190 Gb/s over distances from 5 to 115 m at 230 GHz is demonstrated [7]. However, the processing of baseband signals in these works are relied on offline DSP. The commercially mature digital coherent optical module (DCO) is a promising solution to realize ultra-high data rate real-time communication. At 300 GHz, the real-time transmission of 100 Gb/s net capacity over two fiber links by a pure electronic THz wireless link over 0.5 m distance is demonstrated using DCO modules [8]. In our previous work [9,10], a real-time single-channel 103.125 Gb/s net data rete fiber-THz-fiber system is realized base on hybrid optoelectronic down-conversion at 360 GHz~430 GHz over 0.6 m wireless distance. Still, six manual polarization controllers (PCs) and two individual intensity-modulators are required in the transceiver.

In this work, we propose and show a real-time dual-channel fiber-THz-fiber 2 × 2 multiple-input multiple-output (MIMO) seamless integration system with a record net data rate of 2 × 103.125 Gb/s at 385 GHz and 435 GHz using commercial DCO modules over two spans of 20 km SSMF and 3 m wireless link without using THz power amplifier. To simplify the system architecture, one integrated dual-polarization modulator is used for optical signals remodulation. The performance comparison between single-channel case and dual-channel case are extensively studied.

## Experimental setup

The experimental setup of our real-time dual-channel transparent fiber-THz-fiber 2 × 2 MIMO transmission system based on optical-to-THz (O/T) and THz-to-optical (T/O) conversion is shown in Fig. 1. Four CFP2-DCO modules are used for optical baseband signals real-time processing. The double-carrier frequency of channel 1 (Ch1) and channel 2 (Ch2) is fixed at 193.5 THz and 193.55 THz with 3 dBm output optical power, respectively. The frequency space between Ch1 and Ch2 meet 50 GHz ITU-T grid. Each CFP2-DCO module works at 100 GbE mode, and 31.379 GBaud DP-QPSK modulated optical baseband signal with a roll-off factor of 0.2 is generated. The optical spectrum of the coupled Ch1 and Ch2 is shown in Fig. 2(a).

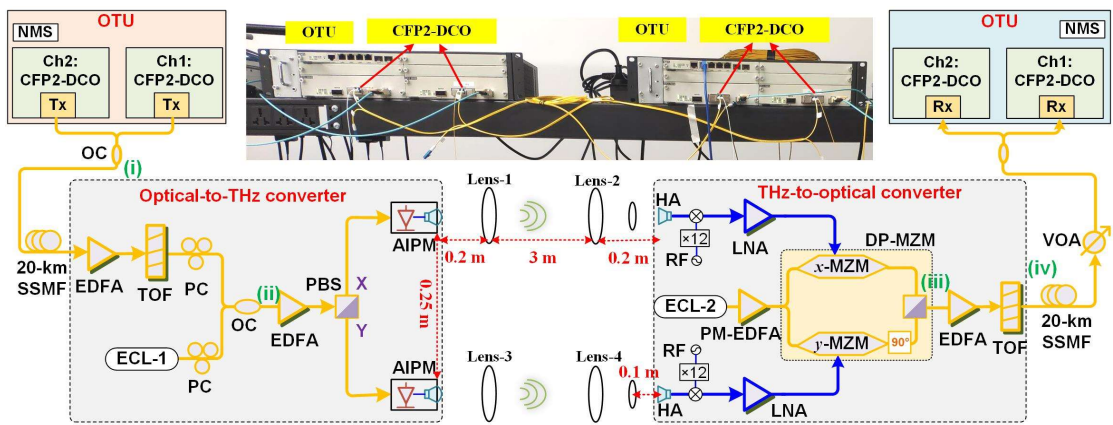

**Fig. 1:** Experimental setup of real-time dual-channel transparent fiber-THz-fiber 2×2 MIMO transmission system.

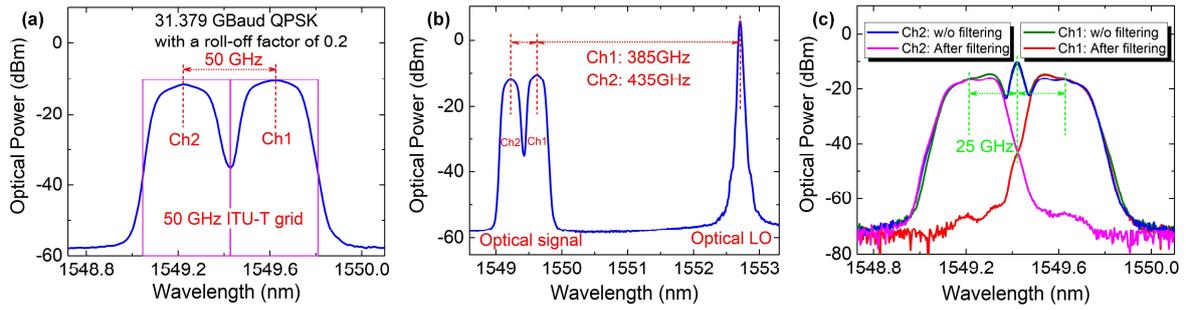

**Fig. 2:** Measured optical spectra at 0.03-nm resolution: (a) the baseband optical signals of dual-channels after OTU;(b) the optical signals and optical LO after optical coupler;(c) remodulation optical signals without and with filtering.

At the O/T converter, over 20 km SSMF, the optical signals with 10.6 dBm optical power and the ECL-1 with 13.5 dBm optical power operating at 193.115 THz as an optical local oscillator (LO) are coupled, and then amplified by an EDFA to effectively drive antenna-integrated photomixer modules (AIPMs) based on uni-travelling-carrier photodiode (UTC-PD). The optical spectrum of the coupled optical signals and optical LO are shown in Fig. 3(b). The frequency space between Ch1, Ch2 and LO is fixed at 385 GHz and 435 GHz, respectively. A polarization beam splitter (PBS) is used to polarization diversity, and X- and Y-polarization components are up-converted by AIPMs to two THz-wave wireless signals at 385 GHz and 435 GHz corresponding to Ch1 and Ch2. The AIPMs are polarization sensitive, and two PCs are necessary to adjust the incident polarization direction to maximize output power from AIPMs. Then, the THz-wave signals are delivered over a 3 m 2×2 MIMO wireless transmission link. Three pairs of lenses are used to focus the wireless THz-wave to maximize the received THz-wave signal power.

At the T/O converter, hybrid optoelectronic down-conversion is used for T/O conversion in order to reduce the carrier frequency and the bandwidth requirement for the modulators. For X- and Y-polarization THz-wave wireless signals, two identical THz receivers are driven by electronic LO sources to implement analog down conversion, and each consists of a mixer, a ×12 frequency multiplier chain and an amplifier. The intermediate frequency (IF) signal bandwidth of THz receivers is 40 GHz. Note that, subject to the availability of the THz receivers, Ch1 and Ch2 is separately measured, but the total 2 × 100 GbE line rate from transmitter is consistent all the time. The sinusoidal LO sources for Ch1 and Ch2 are set 30 GHz and 38.333 GHz, respectively. Hence the frequency of IF signals corresponding to Ch1 and Ch2 are same 385 − 30 × 12 = 25 GHz and 38.3333 × 12 − 435 = 25 GHz. Then, the down-converted X- and Y-polarization IF signals at 25 GHz are boosted by electrical low-noise amplifiers (LNAs) to drive one integrated dual-polarization Mach-Zehnder Modulator (DP-MZM) with 3 dB bandwidth of 35 GHz and 1.8 V half-wave voltage, which operating at optical-carrier-suppression (OCS) point. Thus, four manual PCs are saved. The ECL-2 as the optical carrier input of the DP-MZM working at 193.525 THz, and the optical power after PM-EDFA is 19 dBm. Fig. 2(c) show the measured spectra of optical signals remodulation without and with filtering for Ch1 and Ch2, respectively. For Ch1, another TOF is set to filter out the lower sideband and the central optical carrier as well as the ASE noise, only leaving the upper sideband. Similarly, for Ch2, TOF is set to filter out the upper sideband and optical carrier, leaving the lower sideband. The obtained optical baseband signal is delivered over the second span of 20 km SSMF, and then received by CFP2-DCO modules corresponding to Ch1 or Ch2. A variable optical attenuator (VOA) is used to measure optical signal to noise ratio (OSNR) of the receiver. Figs. 2(a)-(c) are corresponding to the test points (i)-(iv) in Fig. 1.

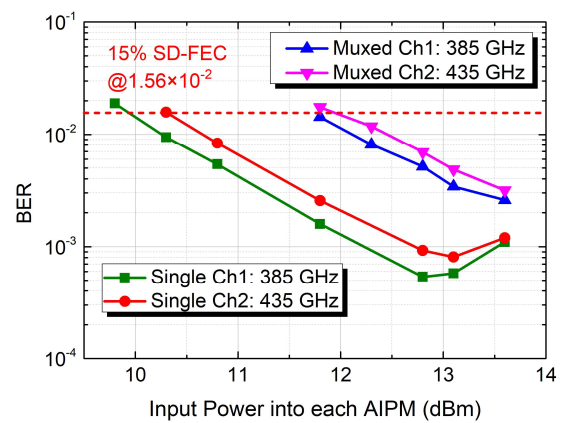

**Fig. 3:** BER versus input power into each AIPM for single channel case and dual-channel case.

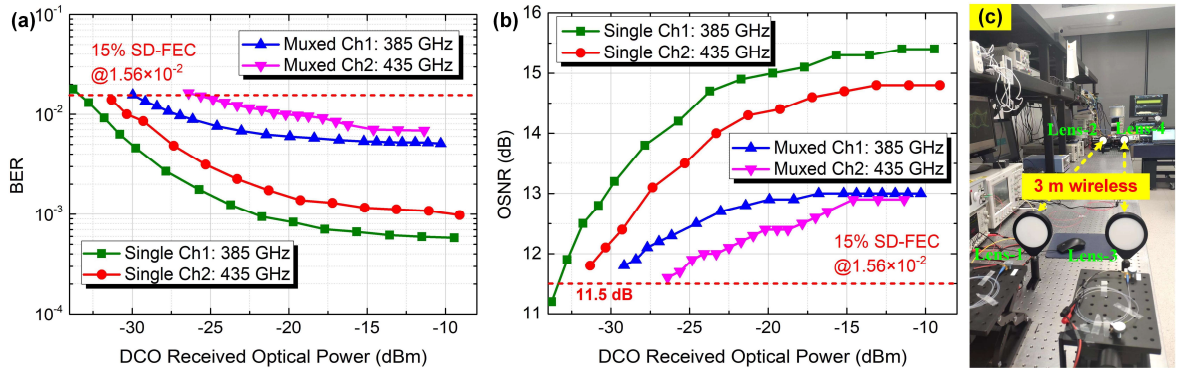

**Fig. 4:** (a) BER and (b) OSNR versus receive optical power (ROP) of each CFP2-DCO module for single channel case and dual-channel case at 385 GHz and 435 GHz, respectively. (c) Photo of 3 m wireless link.

### Results
The parameters, including operating mode, wavelength, output optical power, pre-BER and so on, can be set and monitored by the network management system (NMS). Each CFP2-DCO module can support 15% soft-decision forward-error-correction (SD-FEC) for pre-FEC BER of $1.56\times10^{-2}$ and post-FEC BER$<10^{-15}$. The dual-channel 31.379 GBaud (125.516 Gbps) DP-QPSK signals can provide 2×103.125 Gbps net capacity for 2 × 100 GbE clients.

Fig. 3 shows the BER versus the input power into each AIPM for single channel case and dual-channel case at 385 GHz and 435 GHz over two spans of 20 km SSMF and 3m wireless distance, respectively. For single channel case, the BER performance begins to deteriorate over 13.1 dBm because the power of AIPMs are saturated. About 0.5 dB BER gain is achieved at 15% SD-FEC threshold for Ch1 at 385 GHz compared with Ch2 at 485 GHz. In order to avoid damaging AIPMs, the input optical power remains below 13.8 dBm in the dual-channel case. The BER performance of Ch1 and Ch2 is similar, and there is no power saturation phenomenon since dual-channel multiplexing reduce the average power of each channel. At 15% SD-FEC threshold, there are around 2 dB optical power penalty at 385 GHz and 435 GHz for dual-channel case compared with single channel case.

Then, we evaluate the BER and OSNR performance versus received optical power (ROP) of each CFP2-DCO module over two spans of 20 km SSMF and 3 m wireless distance link. Figs. 4(a) and (b) show the BER and OSNR versus ROP for single channel case and dual-channel case at 385 GHz and 435 GHz, respectively. With increasing of ROP, BER and OSNR are gradually stable. In Fig. 4(a), we can observe that there are around 5 dB optical power penalty at 385 GHz and 435 GHz for dual-channel case compared with single channel case at 15% SD-FEC limit. In Fig. 4(b), the OSNR lower bound is around 11.5 dB corresponding to 15% SD-FEC threshold. Compared with single channel case, there are about 5 dB OSNR penalty at 385 GHz and 435 GHz for dual-channel case. Fig. 4(c) shows the photo of 3 m wireless link. This demonstrated fiber-THz-fiber seamless system can potentially support tens of users for bandwidth-consuming services, such as 3D holographic, metaverse, and 8K/10K video.

### Conclusions
A real-time dual-channel fiber-THz-fiber 2 × 2 MIMO seamless transmission system with a record net data rate of 2 × 103.125 Gbps under 15% SD-FEC is experimentally demonstrated at 385 GHz and 435 GHz THz band using the commercial DCO modules over two spans of 20 km SSMF and 3 m wireless link. O/T conversion at the transmitter is based on photomixing without THz power amplifier, T/O conversion at the receiver relies on hybrid optoelectronic down-conversion using commercial devices. One integrated DP-MZM and only two PCs are used to simplify the system. It's a promising scheme to meet the demands of future THz seamless integration for low power consumption, low cost and miniaturization. Photonics combined with electronic active devices may lever the THz wireless transmission distance for km-range.

### Acknowledgements
This work was partially supported by National Natural Science Foundation of China (62101121, 62101126), Major Key Project of PCL (PCL2021A01-2), China Postdoctoral Science Foundation (2021M702501), and Key Research and Development Program of Jiangsu Province (BE2020012).